**A multi-band analogue frequency sensor with sub-MHz resolution based on a Vortex Nano-Oscillator**

*Alex S. Jenkins*[*,1], *Lara. San Emeterio Alvarez*[1], *Roberta Dutra*[2], *Rubem L. Sommer*[2], *Paulo P. Freitas*[2], *Ricardo Ferreira*[2]

1- International Iberian Nanotechnology Laboratory, INL, Av. Mestre José Veiga s/n, 4715-330, Braga, Portugal.

2 - Centro Brasileiro de Pesquisas Físicas (CBPF), Rua Dr. Xavier Sigaud 150, Rio de Janeiro 22290-180, Brazil

Abstract

**As the Internet of Things (IoT) becomes more integral to how we live our lives and society becomes ever increasingly connected, the need for efficient nanoscale radio frequency detection is becoming more of a pressing issue. The commonly used technique of a superheterodyne conversion of the incoming signal to an intermediate frequency (IF), requires a relatively large electronic circuit, including a local oscillator, mixer, various filters and a rectifier (i.e. diode). In order to radically simplify this process, we propose a spintronics-based nanometric single device whose resistance varies linearly as a function of the incoming frequency. This analogue frequency sensor allows a direct measurement of the incoming frequency and can work over many frequency bands and could replace a significantly larger conventional electronics circuit, thus reducing the footprint and potentially the energy cost. Three types of operation of the frequency sensor are demonstrated: a HF/VHF sensor based on a subharmonic modulation working across 9 different bands with 1-10 MHz bandwidth and 200 kHz resolution, a VHF/UHF sensor based on direct gyrotropic excitation operating between 150-225MHz with 250 kHz resolution and a UHF/SHF sensor based on indirect spin wave excitation operating between 2-4GHz with a 2.5 MHz resolution.**

Spintronics has emerged as an exciting field for generating new paradigms for emerging technologies, and at the forefront of this is the magnetic tunnel junction (MTJ). Comprised of two magnetic layers, separated by an insulating layer, a magnetic tunnel junction has been put forward as a candidate for conventional [1] and bio-inspired [2] memories, magnetic hard drive read heads [3], magnetic field sensors [4] and high frequency components [5]–[17]. The latter of these is based on the so-called spin-torque nano-oscillator (STNO), where the dynamic behaviour of the magnetic tunnel junction creates the potential for a range of applications, including as radio frequency generators [5], [6], [10]–[14] and detectors [7]–[9], [16]–[18].

In this paper we present a new and exciting future application for the STNO, as an analogue radio frequency sensor. Analogous to the MTJ-based magnetic field sensor [4], we present a system whose resistance varies linearly with the incoming frequency, which has the potential to radically change processing of high frequency signals.

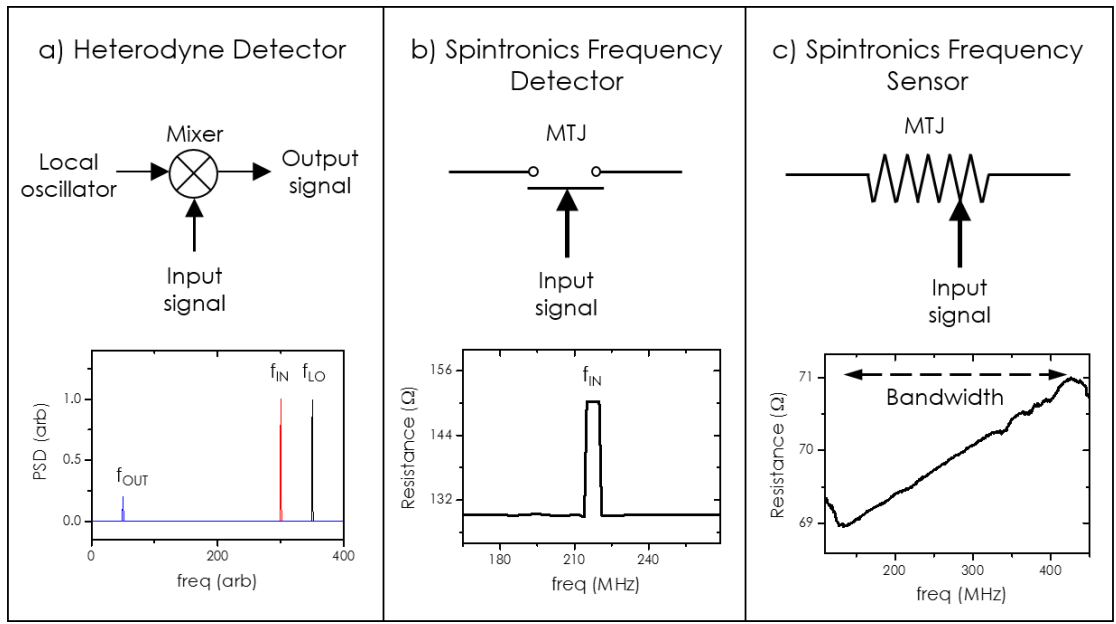

*Figure 1. Schematic demonstration of three different types of frequency detection schemes, a) a conventional heterodyne detector, and a spintronics-based frequency b) detector and c) analogue sensor.*

In Figure 1, the basic working principle of three types of frequency detection are presented. The first, in Figure 1 a), is a heterodyne detector, where the incoming signal ($f_{IN}$) is mixed with the local oscillator ($f_{LO}$), and the resultant intermediate low frequency signal ($f_{IF} = f_{LO} - f_{IN}$) is analysed. Although this commonly used technique works well, it has a large number of components (i.e. local oscillator, mixer, various filters and a rectifying diode), and as such is limited in terms of reduction of footprint and energy consumption, and requires more complex circuitry if integrated with CMOS.

The second detection scheme is a spintronics-based frequency detector shown in Figure 1 b). In this detection scheme, if the incoming signal is within a limited bandwidth around the resonant frequency of the system, a change in the measured voltage or resistance is observed. This change has been demonstrated in MTJs in two ways: firstly by a voltage rectification effect called the spin torque diode effect, first presented in ref [7], and secondly as a variation of the resistance related to a change in the magnetic state, proposed in ref [16], relating to a transition from a magnetic vortex to a quasi-uniform magnetisation. In the latter of these two detector schemes, once the transition between the vortex and the quasi-uniform state is complete, the free layer remains in the quasi-uniform state indefinitely as long as the experimental conditions remain unchanged.

This type of spintronics frequency detector is essentially a frequency dependent switch, and is an efficient means of detection of an incoming radio frequency signal, but crucially, each device will have a unique working frequency and so, to cover a broad bandwidth, either many individual MTJs are necessary (non-tunable) or the system must be tuned with a variable applied magnetic field (tunable). For the non-tunable frequency detector, a careful engineering of the dimensions is required in order to fine tune the resonance frequencies of each nanodevice, which is a significant challenge to be overcome before practical implementation of a broadband detector is possible. The tunable frequency detector schematic can use a single device to cover a range of frequencies, but it is necessary to apply a substantial variable magnetic field, which makes system integration for potential applications extremely challenging and with a substantial energy cost.

The spintronics-based analogue frequency sensor proposed in this article, presented in Figure 1 c), behaves as a variable resistor, where the relative value of the resistance is determined by the value of the incoming frequency. This sensor has the advantage of operating over a broad range of frequencies, such that a single oscillator can be used to detect many frequencies within a finite bandwidth, reducing the spatial dimensions, operating energy cost and the complexity of the final system. Also the one-to-one nature of the frequency to resistance relationship for the analogue sensor means that substantially higher resolutions could be imagined with a frequency sensor relative to the frequency detector. This is due to the fact that the non-tunable spintronics detector schematic requires many MTJs operating for different incoming frequencies, and as such the minimum resolution is given by the reliability of the nanofabrication process in creating many oscillators with slightly different resonant frequencies. By contrast, the analogue frequency sensor comprises of a single MTJ, and the resolution will be determined by the linearity of the resistance versus frequency, and the noise present in the MTJ.

An additional advantage of the spintronics based frequency detector/sensor over the heterodyne detection scheme is that as the high frequency signal will be rectified into a dc voltage, this allows easy integration with CMOS, which would require only the measurement of a dc voltage.

In this report, we show how vortex-based magnetic tunnel junctions [10], [11] are a prime candidate to operate as analogue frequency sensors. We demonstrate how the linear behaviour is related to time-dependent periodic transitions between the vortex to quasi-uniform states, presented here for the first time, and show how these nano-devices can be used to demonstrate three separate types of linear resistance behaviour: a sub-harmonic modulation (Figure 2), a direct gyrotropic resonant excitation (Figure 3) and an indirect excitation due to the coupling between the higher order spin wave modes and the gyrotropic mode (Figure 4).

In Figure 2 we discuss in more detail the physical mechanism behind the frequency dependent variable resistor presented in Figure 1 c). The system under investigation is a magnetic tunnel junction, where for a particular geometry the free layer has a magnetic vortex as the magnetic ground state, as shown in Figure 2 a). The magnetic vortex is a magnetic state where the in-plane magnetisation curls around a central out-of-plane region, known as the vortex core. It is important to note that the following results have been measured for a range of different diameters of nanopillars (300-1000 nm) and similar effects were observed for all of the devices, with the mode frequencies changing as a function of diameter. In Figure 2, Figure 3 and Figure 4, three different devices were selected in order to demonstrate each of the separate effects in more detail (i.e. sub-harmonic, resonant and indirect spin wave behaviour), but all three effects can be observed in a single device, presented in the supplementary information.

In order to excite the magnetic vortex in the free layer of the MTJ, an alternating current (ac) is applied to a 3µm wide field line which is integrated above the MTJ. When the ac current is greater than zero, the resultant magnetic field felt by the free layer is positive, and the free layer will be in the magnetic vortex state, shown in Figure 2 a). When the sign of the ac current changes, the magnetic field generated by the field line will be negative and the vortex core is expelled and the system enters the quasi-uniform magnetisation state. The free layer will constantly alternate between the two magnetisation states as a function of time, depending on the sign of the ac current.

As well as the ac current, a direct current (dc) is also applied to the integrated field line. By tuning the dc current, we can precisely control which magnetic state (i.e. vortex, quasi-uniform parallel (P) or quasi-uniform anti-parallel (AP)) of the free layer is present for the positive/negative part of the alternating current, which is discussed further in the supplementary information.

In this report the transitions between the vortex and the quasi-uniform state are shown via electrical measurements. Due to the tunnelling magnetoresistance (TMR) in the MTJ, the resistance of the vortex state will be quite different to that of the quasi-uniform state. In Figure 2, the static resistance measured with a conventional source-meter (Figure 2 c)) and the time-varying resistance in the time domain measured with a single-shot oscilloscope (Figure 2 b)) are presented. The time domain data was measured in the rf channel of a bias tee, and as such the voltage is displayed in arbitrary units.

In Figure 2 b) it can be clearly seen that the free layer continuously alternates between the static quasi-uniform state (shaded region) and the dynamic vortex state. The peaks observed in the vortex state correspond to the vortex core gyrating at the resonant gyrotropic frequency, around $f_g \sim 150$ MHz. The number of peaks correspond to the number of distinct gyrations performed by the vortex core when the free layer is in the vortex state.

The number of gyrations, n, will depend on the frequency of the ac current, so that for very low frequencies (i.e. $f_{source}$ = 10 MHz), the vortex will gyrate n = 10 times before the free layer returns to the quasi-uniform state, whereas at relatively higher frequencies (i.e. $f_{source}$ = 50 MHz) the core can only gyrate n = 3 times before the free layer subsequently returns to the quasi-uniform state. This leads to the separation into distinct bands located at a sub-harmonic of the resonant gyrotropic frequency, $f_g$. In Figure 2 c), there are 9 separate bands, corresponding to the sub-harmonics f/2 to f/10. This separation into bands can be illustrated by looking at the time traces at specific frequencies, shown in Figure 2 b). At position 1 ($f_{source}$ = 38.0 MHz), the free layer transitions between the quasi-uniform state and the vortex state. When in the vortex state, there is sufficient time for four clear gyrations before the free layer returns to the quasi-uniform magnetisation state. When the frequency is increased to position 2 ($f_{source}$ = 41.4 MHz), there is a finite probability the vortex will gyrate either three or four times. If the vortex gyrates four times then it will subsequently spend less time in the quasi-uniform state ($t_{QU}$ = 9.9 ns), whereas if the vortex gyrates only three times, then it will spend more time in the quasi-uniform state ($t_{QU}$ =12.7 ns). At position 3 ($f_{source}$ = 46.2 MHz), there is only sufficient time for the vortex to gyrate three times.

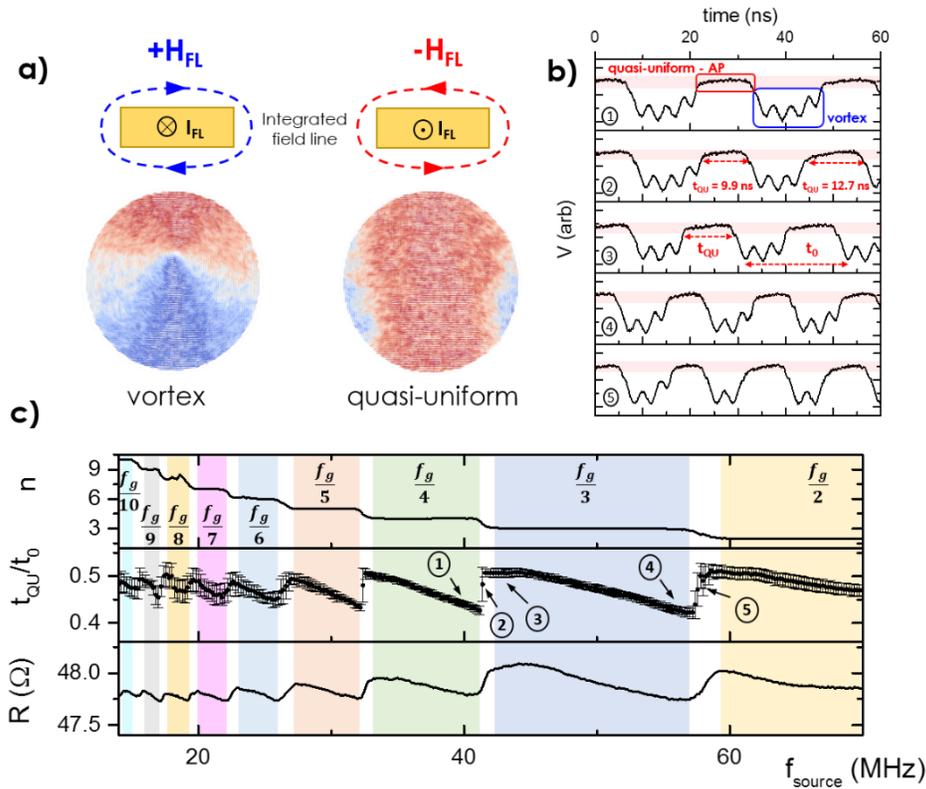

Figure 2 – demonstration of the frequency sensor via low frequency modulation between the vortex and the quasi-uniform magnetisation states, for a nanopillar of 300 nm diameter. a) The vortex and the quasi-uniform magnetisation state are shown for different signs of the current applied in the adjacent integrated field line. b) The voltage measured across the MTJ as a function of time at five different frequencies and c) the number of peaks, n, the average ratio of the time spent in the quasi-uniform state, $t_{QU}/t_0$, and the resistance as measured with a source-meter as a function of the frequency of the current applied in the field line.

The number of gyrations and the frequency of the source will determine the time spent in the quasi-uniform state, $t_{QU}$, and the total time for one period, $t_0$. The average ratio of this, plotted in Figure 2 c), is a measure of how long each cycle the free layer is in the quasi-uniform state. It can be seen that within a band centred on a sub-harmonic, the ratio depends linearly on the applied frequency. This can be seen by contrasting positions 3 and 4 ($f_{source}$ = 46.2 and 54.6 MHz, respectively). In both cases there is only sufficient time for three gyrations, however, it can be seen that the time spent in the quasi-uniform state depends upon the applied frequency ($t_{QU}$ = 9.6 ns and 5.9 ns for position 3 and 4 respectively).

The more time per period the free layer spends in the quasi-uniform state, which in this case is the relatively high resistance anti-parallel state, the closer the average overall magnetisation will be to that of the anti-parallel state, and therefore the larger the time averaged resistance will be. The linear time averaged resistance measured with a source-meter is shown in Figure 2 c) can be easily understood as being directly linked to the amount of time spent in the quasi-uniform AP state.

Once the frequency continues to increase to position 5 ($f_{source}$ = 58.0 MHz), the time spent in the vortex state becomes shorter and as such there is now a finite probability of the vortex gyrating three times or twice, similar to the case of position 2.

The sign of the linear slope observed in Figure 2 depends on whether the free layer transitions from vortex to the anti-parallel state (i.e. negative slope) or vortex to the parallel state (i.e. positive slope). The resultant state after the vortex has been removed by the field depends upon any in-plane magnetic fields and can therefore be easily controlled by a dc current in the field line, see supplementary information.

The linear resistance behaviour originating from the amount of time spent in the quasi-uniform state is further explored in Figure 3. In Figure 2, a low frequency in-plane field induced modulation between the quasi-uniform state and the vortex state was demonstrated, but in Figure 3 the in-plane magnetic field is now applied at frequencies equivalent to the gyrotropic frequency of the vortex, which for the device under testing was around 200 MHz. Similar to Figure 2, in Figure 3 a), the resistance of the MTJ is plotted as a function of the frequency of the alternating current in the field line. The resistance can be seen to vary linearly as a function of the frequency. As with Figure 2, the corresponding time domain data is plotted at various points to show the transitions between the quasi-uniform and vortex states. In this case the time domain data is slightly less intuitive, as there is only sufficient time for one gyration before the vortex is expelled, but none the less, at position 1 ($f_{source}$ = 140 MHz), the free layer can be seen to oscillate between the quasi-uniform state (shaded region) and the vortex state. When the frequency is increased, to positions 2 and 3 ($f_{source}$ = 188 MHz and 220 MHz, respectively), the vortex spends an increasingly larger amount of time in the vortex state and less time in the quasi-uniform state.

The average time spent in the quasi-uniform state as a ratio of the time for a single period ($t_{QU}/t_0$) is plotted in the inset in Figure 3 a). As the frequency increases, the free layer spends less time per period in the quasi-uniform state, and as such the resistance decreases.

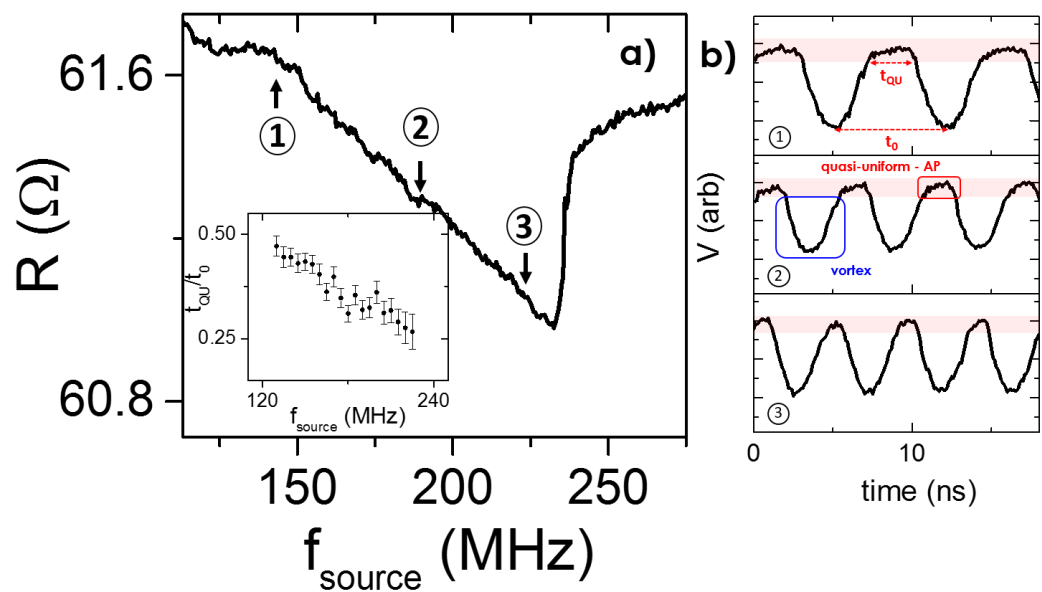

Figure 3- demonstration of the frequency sensor via direct resonant excitation of the gyrotropic mode in a nanopillar with a 600 nm diameter. a) The resistance of the MTJ as a function of the high frequency

*current applied to the field line (inset the average ratio of the amount of time spent in the quasi-uniform state, $t_{QU}$, for a single period, $t_0$). b) The measured voltage showing the system transition between the quasi-uniform state and the vortex state at positions 1, 2 and 3 ($f_{source}$ = 140, 188 and 220 MHz respectively)*

When the frequency becomes larger than the resonant gyrotropic frequency (i.e. $f_{source}$ > 230 MHz), the magnetisation of the free layer no longer transitions constantly between the two states but rather remains in a single magnetisation state, in this case the quasi-uniform state.

The mechanism related to the linear behaviour demonstrated in Figure 2 and Figure 3 is related to a systematic transition of the magnetisation in the free layer between the vortex and the quasi-uniform state, and the linear behaviour is directly related to the amount of time spent in the quasi-uniform state. In Figure 4, another potential mechanism for the generation of the linear resistance versus frequency dependence necessary for the proposed analogue frequency sensor is presented.

In Figure 4, a higher order spin wave mode, present in vortices confined in nanopillars, is excited which indirectly leads to a gyrotropic motion of the vortex core due to the inter-mode coupling created due to the spatial overlap of the modes [19]–[21]. These spin wave modes have been previously reported via optical methods, where switching of the vortex core was achieved with a similar spin wave-gyrotropic mode coupling [21]. It is not within the scope of this study to present a detailed investigation of the nature of these higher order modes, but rather to demonstrate how these modes can be utilised as a frequency sensor.

Due to the coupling between the spin wave modes and the gyrotropic mode, the excitation of the higher order modes can result in a gyrotropic motion of the vortex core. If this excitation is sufficiently strong, then the core will be expelled and the free layer will enter the quasi-uniform state. The strength of this coupling, and therefore the probability of the free layer entering the quasi-uniform magnetisation state, will depend upon the frequency of the ac current. This effect is demonstrated in Figure 4 a), where a linear dependence in the resistance of the MTJ is observed as a function of the frequency of the ac current in the field line. When the voltage is measured in the time domain, as shown in Figure 4 b), the system can in fact be seen to switch between the vortex state and the quasi-uniform state (which in this case is the parallel orientation relative to the reference layer) as a function of time. The data presented here is heavily smoothed so as to remove the effect of the high frequency leakage current, see supplementary info.

In Figure 4 a), the resistance versus frequency behaviour is displayed for a nanopillar where the natural frequency of the spin wave mode is around 4 GHz. At position 1 ($f_{source}$ = 2.14 GHz), the frequency of the current in the field line is far away from the spin wave mode. A gyrotropic motion of the vortex core can be observed, with only very infrequent expulsions of the vortex core. When the frequency is increased and is closer to the natural frequency of the spin wave mode, i.e. position 2 ($f_{source}$ = 3.44 GHz), the system starts to switch more frequently between the vortex and quasi-uniform state. The time spent in the quasi-uniform state increases as the frequency gets closer to the spin wave mode, i.e. positions 3 and 4, ($f_{source}$ = 3.64 and 3.92 GHz, respectively), until at position 4 where the free layer is mostly in the quasi-uniform state. Unlike in Figure 2 and Figure 3, the linear behaviour observed in Figure 4 is related to the probability of the system being in the quasi-uniform state. In the inset in Figure 4 a), the probability of the system being in the quasi-uniform state, $p_{QU}$, is shown for a $t_{window}$ = 2 μs window, by measuring the total amount of time spent in the shaded region normalised by the window length, i.e.

$p_{QU} = t_{QU}/t_{window}$. As the frequency approaches the resonant frequency of the spin wave, the free layer spends less and less time in the vortex state, due to the core being resonantly expelled, and the probability of the system being in the quasi-uniform state increases.

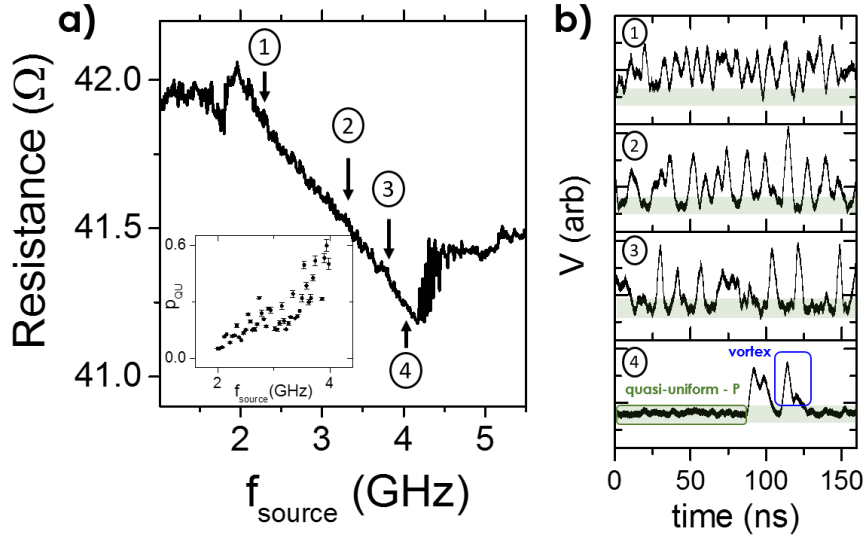

Figure 4 - demonstration of the frequency sensor via indirect excitation of higher order modes of a nanopillar with a 900 nm diameter. a) The resistance of the MTJ as a function of the frequency applied to the field line (inset shows probability of FL being in the quasi-uniform state over a 200 ns window). b) The measured voltage as a function of time showing the quasi-uniform state (shaded region) and vortex state transitions.

Due to the limit of the signal generator used for this experiment, the frequency sensor has not been demonstrated at frequencies larger than 6 GHz, but spin wave modes in confined nanopillars exist across the super high frequency (SHF) band and as such could potentially target new 5g ICT applications.

Having discussed the mechanism for generating a linear resistance dependence necessary for the analogue frequency sensor, in Figure 5 we discuss an important figure of merit, namely the frequency resolution of the sensor. The minimum resolution was acquired by measuring the voltage with a standard voltmeter measuring at 2 ms for 1000 consecutive measurements. The distribution has been calculated, and is plotted in Figure 5, and the minimum resolution is defined as the difference in frequency necessary in order to resolve two normal distributions with a 95% accuracy (i.e. 2σ). The minimum resolution for the modulation and gyrotropic excitations can be seen to be roughly equivalent (200 and 250 kHz, respectively), which is consistent with the fact that in both cases the average time spent in the quasi-uniform state per period is the dominating factor. The minimum resolution for the indirect excitation via the spin wave mode is an order of magnitude larger (2.5 MHz) due to the probabilistic nature of the effect, discussed in Figure 4.

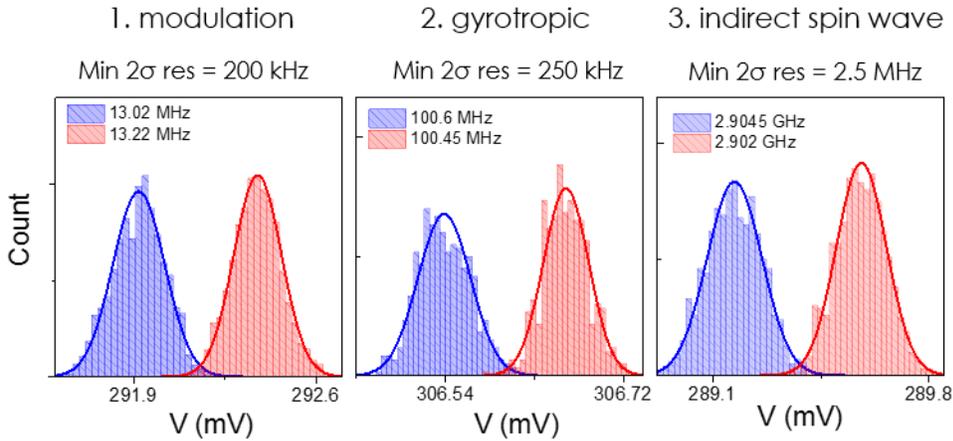

*Figure 5 – Distribution of voltage over 1000 measurements for different applied frequencies, for the modulation, gyrotropic and indirect spin wave mode excitations discussed in Figure 2, Figure 3 and Figure 4 respectively.*

In conclusion, we present a novel type of spintronics-based analogue frequency sensor, using vortex to quasi-uniform transitions in the free layer of a magnetic tunnel junction. We demonstrate three types of frequency sensor: a HF/VHF sensor with 9 bands based on subharmonic modulation with 1-10 MHz bandwidth and 200 kHz resolution, a VHF/UHF sensor based on direct gyrotropic excitation with ~100 MHz bandwidth and 250 kHz resolution and a UHF/SHF sensor based on indirect spin wave excitation with ~1 GHz bandwidth and 2.5 MHz resolution. All three of these sensors can be achieved on a single nanometric device, allowing for reduction in space and energy costs, and as such represent a significant step forward for spintronics-based wireless communications applications.

Methods

The system under investigation is a magnetic tunnel junction based in the stack 5 Ta / 50 CuN / 5 Ta / 50 CuN / 5 Ta / 5 Ru / 6 IrMn / 2.0 CoFe30 / 0.7 Ru / 2.6 CoFe40B20 /MgO wedge 1x80_160% 3kW 600sccm / 2.0 CoFe40B20 / 0.2 Ta / 7 NiFe / 10 Ta / 7 Ru (thicknesses in nanometers) deposited over a 200mm wafer. The MgO layer was deposited as a wedge with variable thickness resulting in an RxA distribution between 1-30 $\Omega\mu m^2$. Similar trends were observed for all values of RxA. The devices were patterned into circular nano-pillars with diameters in the range between 300nm and 1000nm. On top of the pillar, at a distance of ~700nm from the free layer a current line with a width of 3µm made of 300nm thick Al was integrated with the nano-pillars. This current line can generate a local field in the free layer of about 1.5 Oe/mA, up to a maximum field of 150 Oe. The field generated by the field line is collinear with the pinning direction of the antiferromagnet.

The NiFe free layer has a magnetic vortex as the magnetic ground state. The samples were characterised in the dc and high frequency domains. The dc measurements were performed with a standard sourcemeter and the time domain data was taken with a single shot oscilloscope, with the measurements being performed on the ac channel of the bias tee. During the experimental measurements, a perpendicular field is applied. When the perpendicular field is applied it modifies the relative energy levels of the quasi-uniform state and the vortex state, which changes the probability of transitioning between the two states, which in turn changes the linearity. Similar results were observed at zero perpendicular field, but with reduced linearity due to a more stochastic switching between the quasi-uniform and vortex states. This could be addressed by changing the geometry of the nanopillars or thickness of the magnetic materials used in the MTJ. In Figure 2, a 300 nm device was measured at $H_{perp}$ = 0.3 T, $I_{MTJ}$ = 8 mA applied to the MTJ and $I_{FL}$ = -60 mA applied to the field line, with an alternating current with $P_{source}$ = 18 dBm at the source. In Figure 3, a 600 nm device was measured at $H_{perp}$ = 0.6 T, $I_{MTJ}$ = 5 mA, $I_{FL}$ = -20 mA, $P_{source}$ = 16 dBm and in Figure 4 a 900 nm device was measured at $H_{perp}$ = 0.6 T, $I_{MTJ}$ = 8 mA, $I_{FL}$ = 60 mA, $P_{source}$ = 18 dBm.

**Acknowledgments**

The authors gratefully acknowledge the NORTE-01-0145-FEDER-000019 project and funding from MSCA-COFUND-2015-FP-713640.

**Supplementary information**

The resultant state of the free layer of the MTJ for positive and negative alternating current values applied to the field line can be easily controlled by applying an additional dc current. This means that the total field at any moment produced by the field line will be $H_{FL} = H_{FL}(ac) + H_{FL}(dc)$ shown schematically in Figure 6. In Figure 6, the voltage is plotted as a function of time for three different dc currents in the field line. The resultant state after the vortex is expelled depends strongly on the dc component of the $H_{FL}$

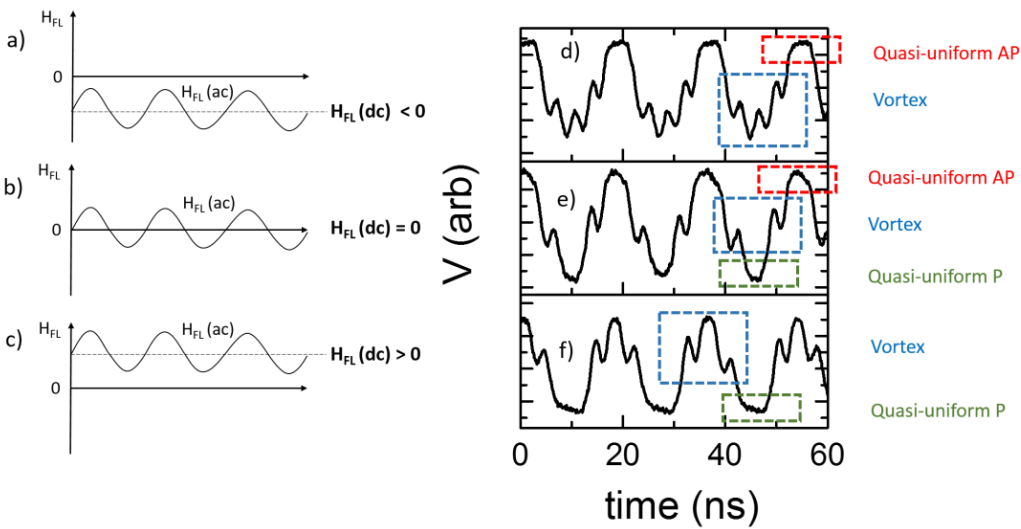

*Figure 6 – Schematic representation of the ac and dc fields applied to the field line and (a) $H_{FL}(dc)<0$, b) $H_{FL}(dc)=0$ and c) $H_{FL}(dc)>0$) the voltage measured as a function of time for the three cases (d) $H_{FL}(dc)<0$, e) $H_{FL}(dc)=0$ and f) $H_{FL}(dc)>0$)*

In Figure 7, three examples of the resonant excitation observed in Figure 3 are shown for three values of $H_{FL}(dc)$ (i.e. positive, zero and negative). The slope of the effect can be seen to depend upon this $H_{FL}(dc)$, as it will determine whether the vortex is expelling into the quasi-uniform parallel (lower resistance relative to the vortex) or quasi-uniform anti-parallel (higher resistance relative to the vortex). As seen in Figure 6, when $H_{FL}(dc)=0$, the vortex expels into both the quasi-uniform parallel and anti-parallel states, which is why a linear slope is not observed because the average resistance is similar to that of the vortex itself.

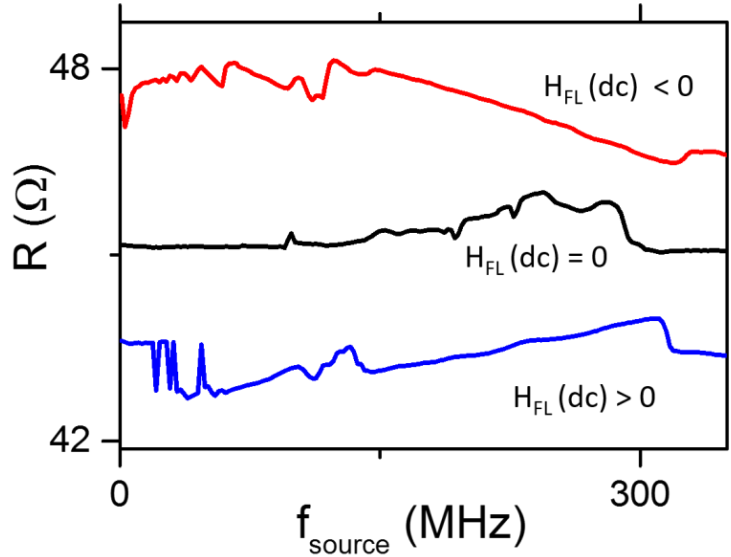

*Figure 7 – The static resistance versus source frequency for three different dc currents applied to the field line, where the vortex transitions between the vortex and quasi-uniform anti-parallel ($H_{FL}(dc)<0$), the vortex and the quasi-uniform parallel state ($H_{FL}(dc)>0$) and the vortex and both the quasi-uniform parallel and anti-parallel ($H_{FL}(dc)=0$).*

In Figure 8, the static resistance has been measured as a function of the source frequency for a single device in order to demonstrate that a single device can exhibit all the three types of excitation (sub-harmonic, resonant and spin wave).

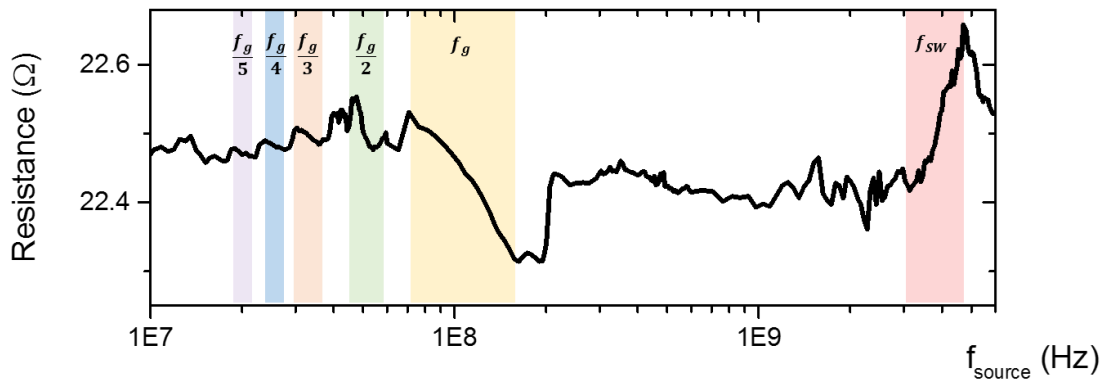

*Figure 8 – measurement of the static resistance versus source frequency for a single device showing the sub-harmonic, resonant gyrotropic and spin wave coupled linear regions for a nanopillar of 900 nm diameter, $H_{perp}$ = 0.5T, $I_{MTJ}$ = 18 mA, $I_{FL}$ = 25 mA and rf power 18 dBm at the source.*

In SIfig3, the time domain data for figure 4 is presented before and after smoothing. The smoothing is performed in order to remove the high frequency leakage current at the source current, and instead allows the lower frequency gyrotropic response to be observed.

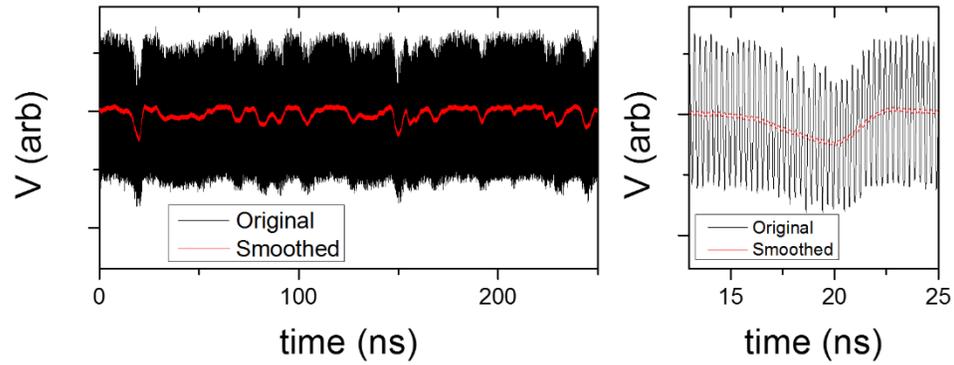

*Figure 9 – The original data and the smoothed data used for Figure 4. The smoothing is performed to remove the high frequency (GHz) source signal and leave the low frequency (MHz) response due to the vortex.*